\documentclass[12pt]{iopart}

\newcommand{\ket}[1]{\ensuremath{\left|{#1}\right\rangle}}

\usepackage{graphicx}

\usepackage{amssymb}

\usepackage{color}

\begin{document}

\title{Validity of resonant two-qubit gates in the ultrastrong coupling regime of circuit QED}

\author{Y. M. Wang$^{1}$, D. Ballester$^{2}$, G. Romero$^{2}$, V. Scarani$^{1,3}$, and E. Solano$^{2,4}$}

\address{$^1$Centre for Quantum Technologies, National University of Singapore, Singapore}
\address{$^2$Departamento de Qu\'imica F\'isica, Universidad del Pa\'is Vasco - Euskal Herriko Unibertsitatea, Apdo. 644, 48080 Bilbao, Spain}
\address{$^3$Department of Physics, National University of Singapore, Singapore}
\address{$^4$IKERBASQUE, Basque Foundation for Science, Alameda Urquijo 36, 48011 Bilbao, Spain}
\ead{vivhappyrom@gmail.com}
\begin{abstract}
We investigate theoretically the performance of two-qubit resonant gates in the crossover from the strong to the ultrastrong coupling (USC) regime of light-matter interaction in circuit QED. Two controlled-PHASE (CPHASE) gate schemes---which work well within the rotating wave-approximation (RWA)---are analyzed while taking into account the effects of counter-rotating terms appearing in the Hamiltonian. Our numerical results show that the fidelity of the gate operation is above 96\% when the ratio between the coupling strength and the resonator frequency, $g/\omega_r$, is of about 10\%. Novel schemes are required in order to implement ultrafast quantum gates when increasing the ratio $g/\omega_r$.
\end{abstract}

\maketitle

\section{Introduction}
Circuit quantum electrodynamics (QED)~\cite{Blais04,Wallraff04,Chiorescu04} has become arguably a prominent solid-state architecture for quantum information processing~\cite{NC}. This technology is built from superconducting qubits that interact with microwave fields in one-dimensional resonators. Its potential applications include single- and two-qubit quantum gates via dispersive~\cite{Majer09,Leek09,Gates,DiCarlo09} or resonant coupling~\cite{Geller,Resonant,Bialczak10,Yamamoto10}, three-qubit gate and entanglement generation~\cite{Fink09,Neeley10,DiCarlo10,Mariantoni11,Stojanovic11,Fedorov11}, as well as novel quantum phenomena in circuit QED such as dynamical Casimir effect~\cite{Wilson11}, and the ultrastrong coupling (USC) regime of light-matter interaction~\cite{Ciuti05,Bourassa09,Niemczyk10,Pol10,Ciuti11}. None of them is accessible in existing cavity QED technology. In particular, the prospects of reaching the USC regime, where the coupling strength $g$ becomes an appreciable fraction of the resonator frequency~$\omega_r$ ($0.1\!\lesssim \!g/\omega_r\!\lesssim \!1$), brings us the possibility of speeding up quantum gate operations at time scales of the order of a few nanoseconds. This will help to overcome decoherence and implement scalable quantum computation. Nonetheless, this comes at the expense of losing the intuitive dynamics of the Jaynes-Cummings (JCM) model. Indeed in the USC regime, the well-known rotating-wave approximation (RWA) fails as the counter-rotating terms become relevant~\cite{Braak}. This means that in order to make possible fast high-fidelity quantum gates, a revision of schemes proposed within the RWA is in order.

In this work, we study two schemes for resonant CPHASE gates in the USC regime, which are designed assuming that the RWA holds. In Sec.~\ref{s2}, we analyze the CPHASE gate scheme proposed in Ref.~\cite{Resonant} considering ratios $g/\omega_r$ in the USC regime. In Sec.~\ref{s3}, we propose a modification of the previous CPHASE gate protocol and investigate how it performs when increasing the ratio $g/\omega_r$. Finally, in Sec.~\ref{s4}, we discuss the physical platform where these protocols can be implemented, and we present our concluding remarks.

\section{Resonant CPHASE gate: scheme I}
\label{s2}
In the computational basis of two qubits $\{|g_1,g_2\rangle ,\, |g_1,e_2\rangle ,\, |e_1,g_2\rangle ,\, |e_1,e_2\rangle\}$, a general CPHASE gate is described by the unitary transformation
\begin{equation}
U_{\rm CPHASE} =
\left(
\begin{array}{cccc}
 1 & 0 & 0 & 0\\
 0 & 1 & 0 & 0\\
 0 & 0 & 1 & 0\\
 0 & 0 & 0 & e^{i \theta}
\end{array}
\right),
\end{equation}
where $\ket{g_i}$ and $\ket{e_i}$~($i = 1,2$) are the ground and excited states of the $i$-th qubit.
In particular, if $\theta = \pi$, the above unitary transformation leads to a controlled $\pi$-phase gate, that together with single-qubit rotations form a set of universal gates for quantum computation~\cite{NC}.

In this Section, we study the performance of the scheme proposed in Ref.~\cite{Resonant} for different values of the interaction strength between qubits and resonator field. This protocol is based on the resonant interaction of three-level superconducting qubits and a single mode of the resonator. In this setup, we assume that the qubit transition frequency can be tuned in order to switch selectively on and off its coupling to the resonator. The logical qubits are encoded into the two lowest energy levels $\ket{g_i},\ket{e_i}$, while the third state $\ket{a_i}$ is used as an auxiliary level.

\begin{table}[t]
\centering
\caption{Operation steps of the protocol I.}
\label{p1_t}
\begin{tabular}[t]{|l|c|c|c|}
\hline
\renewcommand{\arraystretch}{10}
Step & Transition & Coupling & Pulse \\
\hline
\hline
(i) Mapping & $\ket{e_2, 0} \rightarrow -i \ket{g_2, 1}$ & $g_{g_2,e_2}$ & $\pi/2$ \\
\hline
(ii) CPHASE & $ \ket{e_1, 1} \rightarrow -\ket{e_1, 1} $ & $g_{e_1,a_1}$ & $\pi$ \\
\hline
(iii) Back Mapping & $\ket{g_2, 1} \rightarrow   -i \ket{e_2, 0} $ &  $g_{g_2,e_2}$  & $3 \pi/2$ \\
\hline
\end{tabular}
\end{table}

\begin{figure}[b]
\centering
\includegraphics[width=0.5\textwidth]{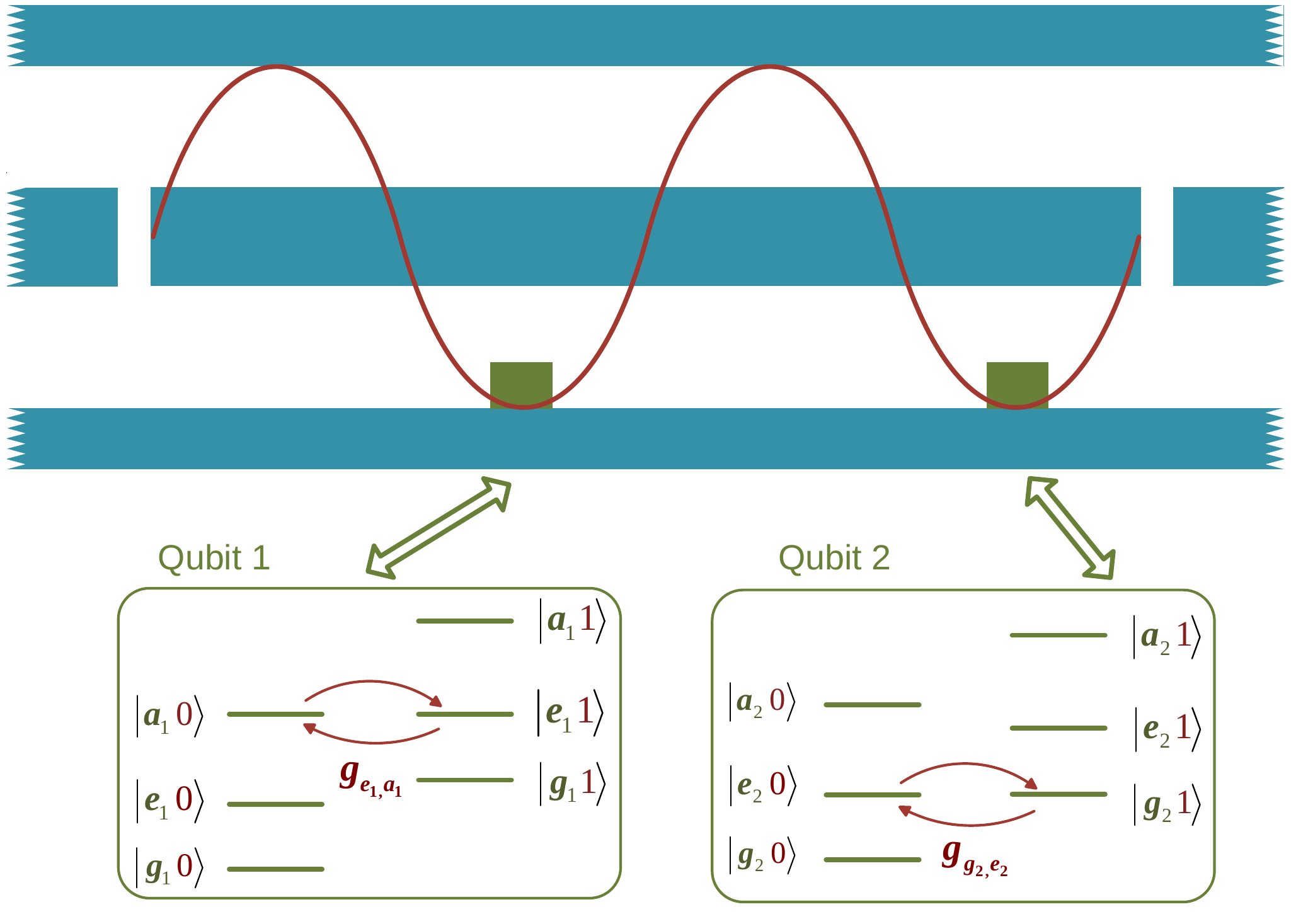}
\caption{Schematic of the protocol I for a resonant CPHASE gate.}
\label{p1_f}
\end{figure}

This first protocol is displayed schematically in Table~\ref{p1_t} and Fig.~\ref{p1_f}. Let us suppose that the initial state of the system reads
\begin{eqnarray}
  |\Psi_{\rm in}\rangle = \Big(b_1 |g_1,g_2\rangle + b_2 |g_1,e_2\rangle + b_3 |e_1,g_2\rangle + b_4 |e_1,e_2\rangle \Big)\otimes\,|0\rangle,
\end{eqnarray}
where $b_i \,(i = 1,\ldots,4)$ are the arbitrary complex coefficients. Within the realm of cavity or circuit QED, it is often the case that the interaction between qubits and resonator field can be described using the JCM, where the coupling strength is small enough so that the RWA is applicable. In such a case, the state evolves to
\begin{eqnarray}
  |\Psi_{\rm out} \rangle= \Big(b_1 |g_1,g_2\rangle + b_2 |g_1,e_2\rangle + b_3 |e_1,g_2\rangle - b_4 |e_1,e_2\rangle \Big)\otimes\,|0\rangle,
  \label{ideal_CPHASE}
\end{eqnarray}
leading to a controlled $\pi$-gate operation.

A natural question is whether this protocol can be extended to higher values of coupling strength, where the interaction lies in the USC regime. This is a relevant question for quantum computation, as it would lead to faster gate operations. We have analyzed the fidelity of the above protocol considering the USC regime, where the RWA breaks down. In this case, The Hamiltonian of the system reads
\begin{eqnarray}
H^q_{\rm NRWA} &=& \sum^{i=1,2}_{j=g,e,a} E_{j_i}\,|j_i\rangle\langle j_i| + \hbar \omega_r \,a^\dag  a \nonumber \\
&+& \hbar g_{e_1,a_1} \sigma_{e_1,a_1}^x \,(a + a^\dag) + \hbar g_{g_2,e_2} \sigma_{g_2,e_2}^x \,(a + a^\dag),
\end{eqnarray}
where $\sigma_{k,l}^x = |l\rangle\langle k| + |k\rangle\langle l|$, $E_{j_i}$ is the energy of the $j$-th level for the $i$-th qubit, and the $g_{e_1,a_1} , \, g_{g_2,e_2}$ are the corresponding coupling strengths.

Taking the state~(\ref{ideal_CPHASE}) as the ideal one to compare with the resulting state of the protocol when including counter-rotating terms in the qubit-resonator interaction, the fidelity can be computed as
\begin{eqnarray}
F&=&{\big|\,\,\langle\Psi_{\rm RWA}\,|\,\Psi_{\rm NRWA}\rangle\,\,\big|}^2 \nonumber \\
&=&{\big|\,\,\langle\Psi_{\rm in}\,|\,U^+_{\rm RWA} U_{\rm NRWA}\,|\,\Psi_{\rm in}\rangle\,\,\big|}^2,
\end{eqnarray}
where $U_{\rm RWA}, \, |\Psi_{\rm RWA}\rangle$ and $U_{\rm NRWA}, \, |\Psi_{\rm NRWA}\rangle$ are the evolution operator and final state in RWA and non-RWA cases, respectively.

\begin{figure}
\centering
\includegraphics[width=0.6\textwidth]{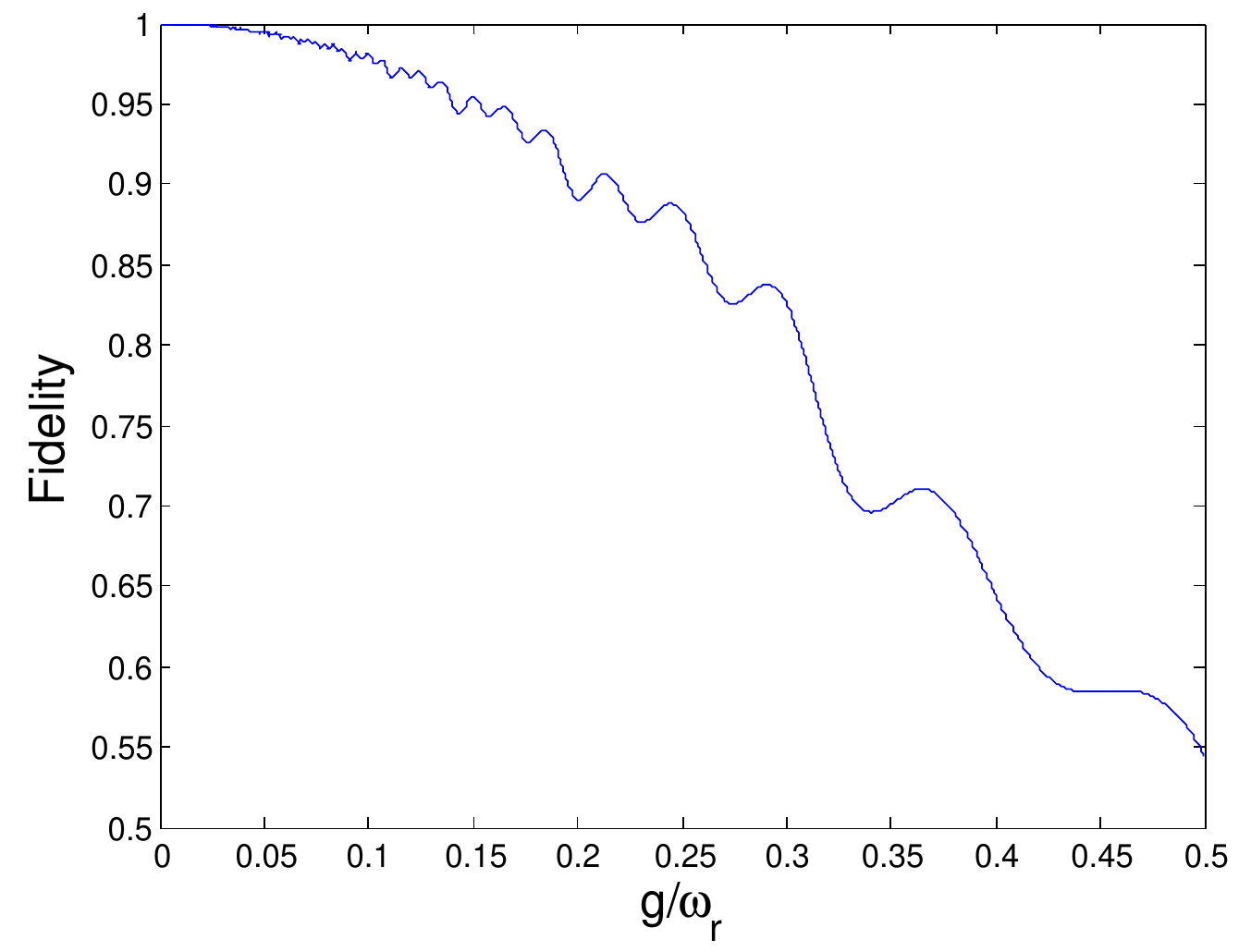}
\caption{Fidelity versus normalized coupling strength $g/\omega_r$ for the CPHASE gate in scheme I with maximally entangled initial state.}
\label{p1_f_f}
\end{figure}

For the sake of simplicity, we assume the coupling strengths between each qubit and the resonator are equal, $g_{e_1,a_1} = g_{g_2,e_2} = g$, and study the case of an initial maximally entangled state
\begin{eqnarray}
  |\Psi'_{\rm in}\rangle = \frac{1}{\sqrt{2}}\,\Big(|g_1,g_2\rangle + |g_1,e_2\rangle + |e_1,g_2\rangle + |e_1,e_2\rangle \Big)\otimes\,|0\rangle.
\end{eqnarray}
The fidelity of the operation as a function of the normalized coupling strength $g/\omega_r$ is shown in Fig.~\ref{p1_f_f}.
Our simulation shows that the fidelity of the gate operation decreases while increasing the coupling strength. For a ratio $g/\omega_r =0.065 $, the fidelity drops below $0.99$, while for $g/\omega_r =0.12 $---which was reached in recent experiments \cite{Niemczyk10,Pol10}---the fidelity goes down to $F \approx 0.968$, and for $g/\omega_r =0.2 $, the fidelity is only $F \approx 0.89$. These results mean that, although this protocol could be still used with state-of-the-art circuit QED technology~\cite{Niemczyk10,Pol10}, its fidelity drops gently as the coupling strength is increased beyond these values.

\section{Resonant CPHASE gate: scheme II}
\label{s3}
\begin{figure}[b]
\centering
\includegraphics[width=0.6\textwidth]{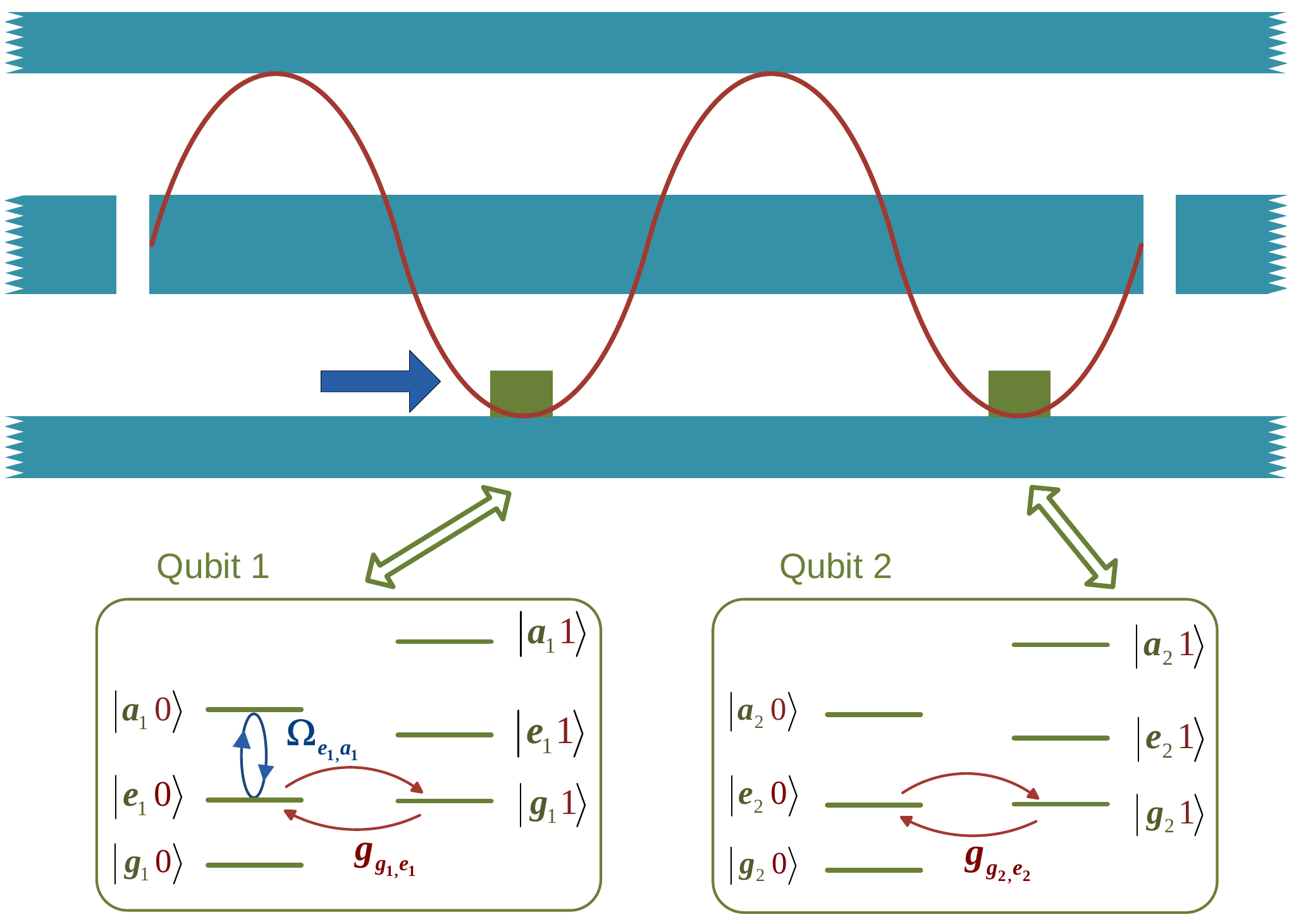}
\caption{Sketch of protocol II for a resonant CPHASE gate.}
\label{p2_f}
\end{figure}

In this Section, we propose an alternative scheme based on a similar configuration, but adding an external microwave field driving the transition $\ket{e_1}\leftrightarrow \ket{a_1}$ of the first qubit, which will assist the gate operation besides the qubit-resonator interaction. The advantage of this modified protocol is that it does not require the accurate adjustment of the qubit transition frequency. The different steps of this scheme are depicted in Table \ref{p2_t} and Fig.~\ref{p2_f}---where a single subsystem has a three-level structure.

\begin{table}[t]
\centering
\caption{Operation steps of Protocol II.}
\label{p2_t}
\begin{centering}
\begin{tabular}[t]{|l|c|c|c|}
\hline
\renewcommand{\arraystretch}{10}
Step & Transition & Coupling & Pulse \\
\hline
\hline
(i) Mapping & $\ket{e_2, 0} \rightarrow -i \ket{g_2, 1}$ & $g_{g_2,e_2}$ & $\pi/2$ \\
\hline
(ii) Rotate qubit 1 & $\ket{e_1} \rightarrow -i \ket{a_1}$ & $\Omega_{e_1,a_1}$ & $\pi/2$ \\
\hline
(iii) CPHASE &  $\ket{g_1, 1} \rightarrow -\ket{g_1, 1}$  & $g_{g_1,e_1}$  & $\pi$ \\
\hline
(iv) Back Rotate qubit 1 & $\ket{a_1} \rightarrow -i \ket{e_1}$ & $\Omega_{e_1,a_1}$ & $ \pi/2$ \\
\hline
(v) Back Mapping & $\ket{g_2, 1} \rightarrow   -i \ket{e_2, 0} $ &  $g_{g_2,e_2}$  & $\pi/2$ \\
\hline
\end{tabular}
\end{centering}
\end{table}

\begin{figure}
\centering
\includegraphics[width=0.5\textwidth]{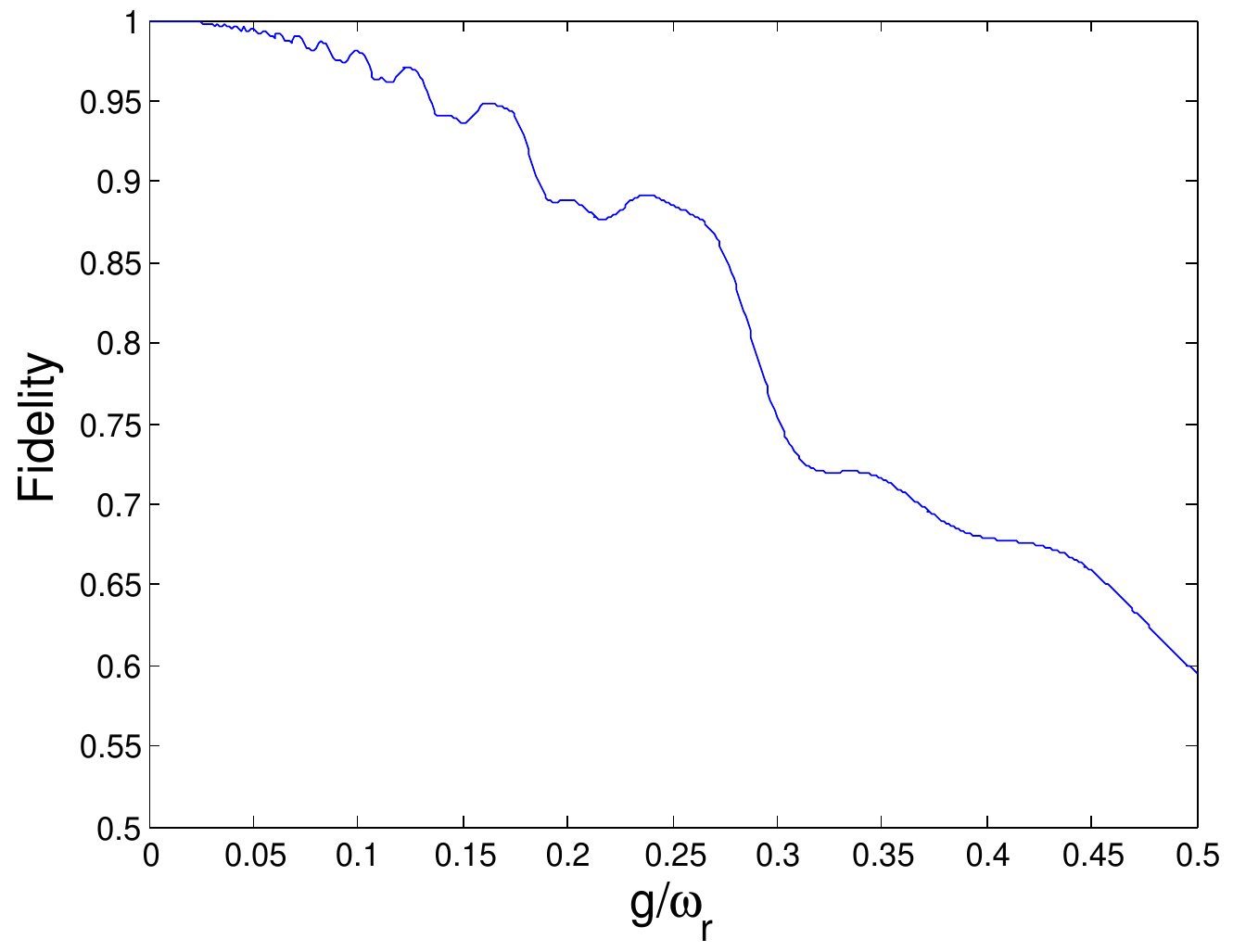}
\caption{Fidelity versus normalized coupling strength $g/\omega_r$ for the CPHASE gate in scheme II with maximally entangled initial state.}
\label{P2_fidelity}
\end{figure}

Within the RWA, protocol II produces the finial state
\begin{eqnarray}
  |\Psi_{\rm out} \rangle= \Big(b_1 |g_1,g_2\rangle + b_2 |g_1,e_2\rangle - b_3 |e_1,g_2\rangle + b_4 |e_1,e_2\rangle \Big)\otimes\,\,|0\rangle,
\end{eqnarray}
that has a $\pi$-phase shift on state $|e_1,g_2\rangle$. To go beyond the RWA in the scheme, we must take into account counter-rotating terms of both qubit-resonator interaction as well as semiclassical model to describe the qubit and classical microwave field interaction. The Hamiltonian for semiclassical model---in the Schr{\"o}dinger picture---without RWA reads
\begin{equation}
  H^c_{\rm NRWA} = E_{e_1} |e_1\rangle\langle e_1| + E_{a_1}|a_1\rangle\langle a_1| + \hbar \Omega_{e_1,a_1} \,\sigma_{e_1,a_1}^x\,\,(e^{i\,\omega_L t}+e^{-i\,\omega_L t}),
\end{equation}
with $\Omega_{e_1,a_1} = d_{e_1,a_1}\,\xi_0/\hbar$ being the Rabi oscillation frequency, and $\omega_L$, $\xi_0 $ being the frequency and amplitude of the classical driving field respectively. Likewise, the corresponding Hamiltonian in rotating frame with respect to qubit frequency one reads
\begin{eqnarray}
  \tilde{H}^c_{\rm NRWA} &=& \Omega_{e_1,a_1} \,\, \sigma_+ \, (e^{-i\,(\omega_L-\omega_{e_1,a_1}) t}+e^{i\,(\omega_L+\omega_{e_1,a_1}) t}) \\ \nonumber
  &+& \Omega_{e_1,a_1} \,\, \sigma_- \, (e^{i\,(\omega_L-\omega_{e_1,a_1}) t}+e^{-i\,(\omega_L+\omega_{e_1,a_1}) t})
\end{eqnarray}
where $\omega_{e_1,a_1} = E_{a_1}- E_{e_1}$ is the transition frequency between levels $|e_1\rangle$ and $|a_1\rangle$.

The wave function in rotating frame can be written as
\begin{equation}
|\Psi'_{(t)}\rangle = C_{e_1}(t)\,|e_1\rangle + C_{a_1}(t)\,|a_1\rangle ,
\end{equation}
where $C_{e_1}(t)$ and $C_{a_1}(t)$ are the complex coefficients for the excited state and auxiliary state, respectively. The corresponding evolution equation for these amplitudes are
\begin{eqnarray}
\dot{C}_{e_1}(t) &=& -i \Omega_{e_1,a_1} (e^{i\,(\omega_L-\omega_{e_1,a_1}) t}+e^{-i\,(\omega_L+\omega_{e_1,a_1}) t})\,C_{a_1}(t) \\
\dot{C}_{a_1}(t) &=& -i \Omega_{e_1,a_1} (e^{-i\,(\omega_L-\omega_{e_1,a_1}) t}+e^{i\,(\omega_L+\omega_{e_1,a_1}) t})\,C_{e_1}(t).
\end{eqnarray}

We have analyzed numerically the protocol fidelity as a function of the ratio $g/\omega_r$. As shown in Fig.~\ref{P2_fidelity}, it is clear that the fidelity decays in a similar fashion as the previous one, as the normalized coupling strength increases, assuming that $\Omega_{e_1,a_1}=g$. Thus this protocol is also unsuitable considering coupling strengths of about $g/\omega_r >0.15$, well within the USC regime.

\section{Discussion and Conclusion}
\label{s4}

Nowadays, the controllability of superconducting devices made of Josephson junctions has led to the implementation of different types of superconducting qubits described by phase, charge, or flux degrees of freedom~\cite{SCQ}. This fast growing technology now allows to access the coherent control of these artificial atoms in a three-level configuration. Leading experiments have shown this ability with transmons~\cite{3level_transmon}, the phase qubit~\cite{Yamamoto10}, and the flux qubit~\cite{3level_flux}. These advances support the implementation of our protocols in various kinds of superconducting circuits.

Besides, the complexity of the protocols is manifested by the degree of difficulty in manipulation, i.e. the number of the pulse sequences needed to perform. Although in our protocols three or four steps are needed to perform the gate, and an external field is used to assist the operation, we do not require direct qubit-qubit interaction as compared with the two-qubit algorithms demonstrated in~Ref.~\cite{DiCarlo09}. In that scheme, the qubit-qubit coupling is realized by a virtual excitation of an intracavity field and this second-order based coupling results in a slower operation. In another experimental realization~\cite{Yamamoto10}, the two-qubit controlled-Z and controlled-NOT gates were realized by making use of an extra capacitor mediating the coupling between qubits.

In conclusion, we have analyzed numerically the performance of those two CPHASE gate schemes in USC regime, which are designed to work ideally in the conventional strong coupling cases. Our results indicate that these two schemes can still work with high fidelity for values of normalized coupling strength of about $g/\omega_r=0.12$. This is the maximum value of coupling ever achieved between quantum light and matter, and was realized in recent experiments using superconducting flux qubits~\cite{Niemczyk10, Pol10}. However, our numerical analysis also shows that given the present prospects to go beyond this coupling strength using similar technology~\cite{Braak,DSC,QSUSC}, there is a need to develop new protocols for quantum gates beyond the RWA, making possible the design of ultrafast quantum gate operations for quantum information processing~\cite{ultrafast}.

\section{Acknowledgements}
We thank Pol Forn-D\'iaz and Frank Deppe for useful discussions.  The authors acknowledge support from the National Research Foundation and the Ministry of Education, Singapore, Juan de la Cierva MICINN Program, Spanish project MICINN FIS2009-12773-C02-01, Basque Government IT472-10, SOLID and CCQED European projects.

\end{document}